\documentstyle[preprint,aps]{revtex}

\begin{document}
\oddsidemargin=-40pt
\baselineskip=18pt
\draft

\preprint{}

\title{van Hove  Singularity effects in  Strongly Correlated Fermions} 

\author{Pinaki Majumdar and H. R. Krishnamurthy$^{\dagger}$} 
\address{ Department of Physics, Indian Institute of Science, Bangalore 
560 012, India. \\
$^{\dagger}$ Also at Jawaharlal Nehru Center 
for Advanced Scientific Research, 
Jakkur,\\ 
Bangalore 560 064, India.}

\date{\today}

\maketitle

\begin{abstract}

We examine the effects of a van Hove singularity (vHs) 
in the density of states (DOS) of a two dimensional, 
$t-t'-U$
Hubbard model  within the local approximation. 
%Even when the interaction $U$ is 
%sufficiently large to drive the half-filled band
%insulating, 
Within our approximation the  
non-interacting Fermi surface  is 
always retained away from 
half-filling; and we find that the vHs in the DOS
survives, and for hole doping is nearly "pinned" to the Fermi
level due to the  enhancement of the
effective mass, $m^*$.
We discuss the electron-hole asymmetry of
this phenomena and its relevance to the  electron and
hole doped cuprates. At optimal (hole)
doping, with the Fermi level at the vHs, we find that 
the single particle damping rate $\Gamma$,
and the resistivity $\rho_{dc}$, scale as $T$ 
for moderately low  temperatures ($T \gtrsim 100$K),
while for the electron doped system
$\Gamma $ and $\rho_{dc} \sim T^2$.

\end{abstract}

\newpage

\narrowtext

The cuprate superconductors are generally regarded 
as doped "Mott" insulators.
The parent state is an antiferromagnetic insulator, with a 
large local moment and a gap $\sim$  O(2ev) \cite{undoped}, 
and cannot be understood within a one particle,
band-theoretic approach. Transport measurements on the
hole doped systems indicate a "carrier density" which scales
with the doping $\delta$.
Naively one might take this to imply 
small hole-like Fermi pockets with area
$\sim \delta$.
Angle resolved photoemission (ARPES) however
reveals \cite{arpesgen} 
a large Fermi Surface (FS)
with area $\sim 1-\delta$
in agreement with the Luttinger theorem, and a shape
roughly consistent with LDA calculations. Thus, 
the one particle spectrum in the hole 
doped system has
band-like features, while the transport is "anomalous" in
its doping and temperature (T) dependence.
On the other hand
there are  systematic differences between the 
hole doped ($p$ type) and electron doped ($n$ type) systems. 
The $p$ type cuprates exhibit
dc resistivity, $\rho_{dc} \propto T$  \cite{ptype} at optimal
doping (maximum $T_c$), have $T_c \sim (40 -
150)K$ and a strongly anisotropic gap $\Delta_{\vec k}$
\cite {ptypegap}
in the superconducting phase.
The $n$ type cuprate  Nd$_{2-\delta}$Ce$_\delta$CuO$_{4-y}$ (NCCO) 
on the other hand shows $\rho_{dc} \propto T^2$, has a
rather low optimal 
$T_c\sim 25K$ and an isotropic gap \cite{NCCO}. 
Analysis of  ARPES data 
provides an interesting clue \cite{arpesgen} to 
what might well be responsible for this
difference between $p$ and $n$ type cuprates.
ARPES clearly reveals \cite{arpesgen} the 
presence of flat bands near $({\pi\over a} ,0)$, {\it i.e} 
a saddle point in the "dispersion",
within 10-20 meV of the
Fermi level $(E_F)$ for $p$ type cuprates. The saddle point
is seen in NCCO also, but
$\sim $ 350 meV below $E_F$, and thus cannot play a
significant role in transport. This suggests a possible
connection between the anomalous $T$ dependence of 
$\rho_{dc}$ and the proximity of the saddle point to $E_F$.

The presence of a
logarithmic van Hove
singularity (vHs) in the density of states (DOS), arising out of
a saddle point in the quasi 2-d dispersion,
had been invoked earlier \cite{vhcomm} as a possible
mechanism for the "high $T_c$" and also as an
explanation for the anomalous transport, but  within a 
weak coupling approach \cite{vhtransp}.
While the presence of a vHs in the bandstructure of a quasi
2-d system is understandable within 
LDA calculations, the survival of this feature
in a strongly correlated system close to the
Mott transition is not, and has not been
demonstrated in any controlled theory so far.
We  address the 
problem of the interplay of bandstructure and strong
correlation effects
in this paper, by studying 
a $t-t'-U$ Hubbard 
model in $d=2$, with a vHs in 
its DOS (at $\sim 0.2$ hole doping) and
with $U$ sufficiently large to 
drive the half-filled ($\delta=0$) system insulating,
using the local approximation.

Within our approximation for the electron self-energy
(detailed below) which retains the non-interacting
FS at $T=0,\delta\neq 0$, we find the
following results $(i)$~The vHs survives
in the DOS of the interacting system, and its shift 
with respect to
$E_F$ with doping is reduced roughly as
$\Delta E(\delta)/m^*(\delta)$
(Fig. 1), where 
$\Delta E(\delta)$ is the shift in the 
noninteracting system, and 
$m^*(\delta)$ is the effective mass enhancement due to
correlation.
$(ii)$~The "dispersion" of the interacting spectra 
(Fig. 2) is
reduced by a factor $\sim 3$ with respect to
the 
noninteracting $t-t'$ band dispersion.
The resulting  "flat-bands" are close to $E_F$
for the optimally doped $p$ type cuprates, and much
farther down for
the $n$ type cuprates. 
$(iii)$~The single particle damping rate $\Gamma$,
(Fig. 3) and
$\rho_{dc}$ (Fig. 4) scale at moderately low
temperatures as $T$  for optimal hole doping, while
they scale as $T^2$ for the electron doped case.
$(iv)$~However, this approximation fails to reproduce the
temperature dependence of the Hall coefficient, $R_H$, 
although the $\delta$ dependence is recovered.

We use the Hamiltonian 
\begin{equation}
\hat{H} = \sum_{\vec k, \sigma}
\left( \epsilon_{\vec k} - \mu \right) 
{c^{\dag}}_{\vec k \sigma}
c_{\vec k \sigma} + U \sum_{i} n_{i\uparrow}n_{i\downarrow}
\end{equation}
where $\epsilon_{\vec k}=-2t(cos(k_x a)+cos(k_y a))
 -4t'(cos(k_x a)cos(k_y a) - 1)$ is the dispersion of the
2-d tight binding square lattice with nearest neighbour
hopping $t$ and next nearest 
neighbour hopping $t'$ (the shift $4t'$ maintains
the band edges at $\pm 4t$).
We set the half-bandwidth $D=1$, $t=D/4$,
$t'=-0.3t$ so that the vHs is located at
the Fermi level for  $\sim 20\%$
hole doping,
and further choose $U/D=4$.
The $t-t'$ model provides the simplified
"universal" 2-d bandstructure of the cuprates, while the
choice $U/D=4$ ensures that the half-filled state is
insulating; the  $t-t'-U$ model can 
thus be regarded as the minimal
model for the cuprates. 

There is no controlled method for studying the doped
phase of the strong correlation problem defined above. We 
adopt an approach which is exact 
for the $t-U$ model in the limit 
when the dimensionality $d
\rightarrow \infty$ \cite{dinf1}, 
but is necessarily an approximation 
for the 2-d problem we want to study;
we assume that the  self-energy and all
the irreducible vertex functions 
are local, {\it i.e} ${\vec k}$ 
independent. In this
approximation the Hubbard model maps on
to a {\it self consistently embedded} Anderson impurity
problem \cite{gkotprb}. A recent method devised by
Kajueter and Kotliar \cite{iptasym} allows an 
interpolative solution of this 
self-consistent impurity 
problem for arbitrary $U$ and $\delta$ at $T=0$,
which is exact in the atomic and band limits,
and makes feasible
the study of the  {\it real frequency} 
single particle spectra. An important
feature of 
this solution is the imposition of the Friedel sum
rule for the impurity problem,
which translates to the Luttinger sum 
rule for the corresponding lattice
problem, ensuring the correct low energy structure
in the spectra, as verified \cite{iptasym} by
comparing with exact results. In an earlier paper \cite{hall}
we have used a natural extension of this method to finite $T$,
in the context of
a symmetric, semicircular DOS 
for the non-interacting band,
to study transport, in particular the
Hall coefficient, in a
doped Mott insulator. We refer the reader to these
two papers \cite{iptasym,hall} for the details
regarding our approximation scheme.
The crucial difference between \cite{hall} 
and the present study is 
in the consistency relation 
for the local propagator,
\begin{equation}
G_{ii}(\omega)
=\int d\epsilon \rho_0 (\epsilon) 
/(\omega + \mu -\epsilon -\Sigma(\omega)).
\end{equation}
We now use a $\rho_0 (\epsilon)$ 
which closely mimics the features of the 
DOS of the 2-d $t-t'$ band, namely the analytic form  
$\rho_0 (\epsilon) = {\rm A}_{\rm N}
{\rm ln}((\epsilon -\epsilon_{vH})^2
+ \eta^2)$, with $\vert \epsilon \vert < D$,
where $\epsilon_{vH} = -0.3D$ is the
location of the vHs
(as in the actual $t-t'$ DOS),
 $\eta=10^{-2} D$ 
regulates the divergence, and
${\rm A}_{\rm N}$ normalises the DOS.
The regularisation makes the numerics simpler, and
in actual systems can be thought of as  arising from
weak interplanar coupling or impurity scattering
effects.

We have studied the self-consistent problem for five
representative  choices of $\delta$; 0.08~(underdoped),
0.20~("optimally" doped), 0.40~(overdoped), and
-~0.20 and -~0.40~(electron doped), for $T \sim 0.01D - 0.1D$.
Since the irreducible vertex functions are 
assumed to be ${\vec k}$ independent, 
vertex corrections to the conductivity are
absent, and the conductivity
$\sigma_{\alpha\beta}$ can be calculated 
from the converged solutions for the
self energy $\Sigma (\omega,T)$. 
The dc conductivity is calculated as \cite{transpcal}
\begin{equation}
\sigma_{xx}(T) =c_{xx} 
\int d\omega ({-\partial f \over \partial \omega})
{1 \over N} \sum_{{\vec k},\sigma}
v^2_x
A^2 ({\vec k},\omega) 
\end{equation}
where 
$c_{xx} = e^2\pi/(2 \hbar a_0)$,$a_0$ is the lattice 
spacing,
$f\equiv[exp(\omega/T) + 1]^{-1}$,
$v_\alpha = \partial{\epsilon_{\vec k}}/\partial k_\alpha$,
and $A( {\vec k}, \omega)
\equiv -{1\over\pi}~Im[\omega+~\mu-~\epsilon_{\vec k}
-\Sigma(\omega)]^{-1}$.
The transverse conductivity is given by \cite{transpcal}
\begin{equation}
\sigma_{xy}(T) =B c_{xy} 
\int d\omega ({-\partial f \over \partial \omega})
{1 \over N} \sum_{{\vec k},\sigma}
u_{xy}({\vec k})
A^3 ({\vec k},\omega) 
\end{equation}
where $c_{xy} = 2{\vert e\vert}^3 \pi^2 a_0/(3 \hbar^2)$,
and $u_{xy}({\vec k}) = v^2_x (\partial v_y/\partial k_y) -
v_x v_y (\partial v_x/\partial k_y)$.
The Hall coefficient $R_H$ is given by 
$\sigma_{xy}/B\sigma_{xx}^2$.

We first discuss the "pinning" of the vHs 
in the total DOS with respect to 
variations in doping. One can  parametrise the
spectral function 
for $\omega \rightarrow 0, T \rightarrow 0$ as
$A({\vec k},\omega) \sim 
\Gamma/((m^*\omega +\bar\mu -\epsilon_{\vec k})^2 
+ \pi^2 \Gamma^2)$
where $\Gamma\equiv -(1/\pi) Im \Sigma(\omega=0)$,
$m^*\equiv 1-(\partial \Sigma_R/\partial \omega)_{\omega =0}$
and $\bar\mu\equiv \mu(\delta,T) - 
\Sigma_R(\delta,T, \omega=0)$.
The Luttinger sum rule ensures that  
$ \bar\mu(\delta,T=0)= \mu(\delta,T=0,U=0)=\mu_0 (\delta)$,
the non-interacting band chemical potential.
At sufficiently low temperatures, $\Gamma/D << 1$, 
so for calculating the small $\omega$
DOS
$A({\vec k},\omega)$ can
be approximated as 
$\delta(m^*\omega +\bar\mu -\epsilon_{\vec k})$.
Since the DOS in the interacting system,
$\rho(\omega)$, is given by 
$-(1/\pi)ImG_{ii}(\omega^+)$
eqn [2] reveals that for $\omega \rightarrow 0$,
$\rho(\omega) \simeq \rho_0 (m^*\omega + \bar\mu)$.
For $T\rightarrow 0$, $\bar\mu \rightarrow \mu_0$,
which directly shows that if the non-interacting system
had the vHs at the Fermi level ($\omega=0$), then
it survives at arbitrary $U$ as long as $m^*$ is
finite.
The form of
$\rho(\omega)$ also reveals that close to optimal doping 
the shift of the vHs from the Fermi level 
with doping is reduced by 
a factor of $m^*$ with respect to the 
non-interacting problem ({\it cf.} Fig. 1).
As we will see later, 
it makes a significant difference to the
transport  that the vHs continues to be
close to  $E_F$ for hole doping 
close to optimal doping, while for the electron doped 
($\delta = -0.2$) case the vHs 
in the DOS is  $\sim 0.06D$ below
$E_F$.

The vHs in the  cuprates is inferred from the 
saddle point observed in the ARPES "dispersion curves".
The saddle point and the accompanying  "flat bands" are
ubiquitious across the cuprates \cite{arpesgen},
having been observed in the Bi 2212 \cite{2212pes}
and 2201 \cite{2201pes}, YBCO \cite{123pes}, and
NCCO \cite{nccopes}. 
However,  comparison of LDA 
calculation and ARPES 
data \cite{nccopes}, 
along
the symmetry lines $\Gamma$ ($0,0$) $\rightarrow X$
(${\pi\over a},{\pi\over a}$) 
$\rightarrow G_1$ (${\pi\over a},0$) $\rightarrow
\Gamma$ ($0,0$), 
reveals that the measured 
"bandwidth", particularly along the $X-G_1$
(zone corner - edge
center)  direction is reduced by a factor of
2 with respect to the LDA result. 
Fig. 2 shows our results for the "dispersion curves",
as inferred from the peaks in our $A({\vec k},\omega)$,
for $\delta =0.2$ (hole doped case) and $\delta=-0.2$
(electron
doped case). For both cases there is a substantial
reduction of the bandwidth, by factors of $\sim 3$ and
$\sim 4$
respectively. In the optimally doped case, the "flat band"
is at $E_F$ and disperses less than $\sim 10$ meV along
$G_1-\Gamma$. In
the electron doped case the "flat band" is $\sim 60$ meV 
below $E_F$ and disperses much more (along $G_1-\Gamma$).
Also worth noting is the asymmetry in the narrowing 
of the dispersion relations for electron and hole
excitations. For the hole doped case, the hole dispersion 
is narrowed more than the electron dispersion as the 
former correspond to the lower Hubbard band which 
contains the Fermi level. In the electron doped case,
the structure is reversed, as the Fermi level lies in 
the upper Hubbard band \cite{footnote}

The singular structure in the DOS 
close to $E_F$ has non-trivial
consequences for the $T$ dependence of the damping rate
$\Gamma$ and the resistivity $\rho_{dc}$.
Although $\Sigma(\omega, T)$ 
is "Fermi-liquid like" for  $T, \omega \rightarrow 0$,
with $\Gamma$ and $\rho_{dc}
\sim T^2$, beyond a low $T$ scale 
that seems to be set by the
proximity of the vHs to $E_F$ the $T$ dependence becomes 
"anomalous" as shown in Figs. 3 and 4.
In Fig. 3 we contrast the $\Gamma$ calculated
with the vHs to that for a symmetric, regular DOS,
$\rho_{0}(\epsilon)=(2/{\pi D})\sqrt{1 - 
(\epsilon/D)^2}$. For both the vHs and the regular DOS
$\Gamma$ starts out as $T^2$, but in the vH case, and
for hole doping it rapidly changes over to a quasi-linear
form. In fact in the regime $T \sim 0.01D -
0.08D$,  at
optimal doping $\Gamma \simeq T$.
The same effect is also seen in the $T$ dependence of
$\rho_{dc}$, as shown in Fig. 4. We have not yet carried
out enough systematic studies to pick out the dependence
of the crossover scale on the parameters of the model.

The  weak coupling vHs scenario for
anomalous transport  has been criticised by
Hlubina and Rice \cite{hricecomm} on the basis that
in this scenario $1/\tau$ is strongly ${\vec k}$ 
dependent. It is anomalous ($1/\tau_{\vec k} \sim T,
\epsilon_{\vec k}$ etc)
only for ${\vec k}$ regions close to the vH saddle points
$(\pm{\pi\over a},0),(0,\pm{\pi\over a})$,
and hence their contribution to
$\rho_{dc}$ will be "shorted out" by the ${\vec k}$ 
regions away from these points, which have a normal,
smaller $1/\tau$ ($1/\tau_{\vec k} \sim
T^2,\epsilon^2_{\vec k}$). This criticism clearly
does not apply to the above strong coupling results, 
where $\Gamma$ is uniformly anomalous over the entire
FS. While the T dependence of $\Gamma$ and $\rho_{dc}$
are rather different from those for the regular DOS,
the results for $R_H$ are not dramatically different
\cite{hall},
except that now the $(T=0)$ noninteracting $R_H$ is also
positive for the hole doped cuprates and a change in sign 
in $R_H$ (at low doping, low T) occurs only for the 
electron doped cuprates.
The local approximation 
does not reproduce the $T$ dependence of
$R_H$ observed in the cuprates.

In conclusion, we have shown that even in the context of a 
simple local approximation, the proximity of a vHs to the
Fermi level leads to interesting strong correlation effects
in ARPES and $\rho_{dc}$ that bear many similarities to
the data in electron doped and in  optimally or
over (hole) doped cuprates. Recent ARPES data on
under(hole) doped cuprates have been interpreted
as showing evidence of a non-Luttinger FS, with regions of
the FS near the vH points becoming "gapped"
\cite{nlfs,wenlee,spvhs}.
There is no way one can access such features within
a "local"
approximation, with a ${\vec k}$ independent $\Sigma$.
One can hope that putting back the ${\vec k}$
dependence of $\Sigma$, and of the vertex functions as
"perturbative", "non local" corrections to the above theory
may cure the failings pointed above.

We thank A. Fujimori, D. D. Sarma, Ashish Chainani and D. M. Gaitonde
for discussions, and Alexander Punnoose for a reading of the
manuscript. We thank the SERC, IISc for providing 
computational facilities.

FIG. 1 The interacting DOS at low frequency. 
For the "optimally doped"
$(\delta=0.2)$ case the vHs remains at $E_F$. For 80\% doping
variation the peaks are within a width of $\sim 0.22$. The
corresponding width in the non interacting system is $\sim 0.42$. 

FIG. 2 "Dispersion" for complementary $p$ and $n$ doping; $U=4, 
T=0.01$. The "flat band" is right at $E_F$ for $\delta=0.2$.
The non interacting dispersion goes from  $-1$ to  $1$ for
the same scan.

FIG. 3 Single particle damping rate for $U=4$. Inset; $\Gamma$
for the symmetric, regular DOS.

FIG. 4 Dc resistivity, in units of $e^2\pi/(2 \hbar a_0)$,
for representative doping.


\begin{references}

\bibitem{undoped} D. B. Tanner and T. Timusk in
{\it Physical Properties of High Temperature 
Superconductors III}, Ed. D. M. Ginsberg
(World Scientific, 1992)

\bibitem{arpesgen}see  Z. X. Shen and D. S. Dessau, 
Physics Reports {\bf 253}, 1 (1995), Z. X. 
Shen {\it et al},
Science {\bf 267}, 343 (1995) and references therein.

\bibitem{ptype} Y. Iye in
{\it Physical Properties of High Temperature 
Superconductors III}, Ed. D. M. Ginsberg
(World Scientific, 1992)

\bibitem{ptypegap} D. A. Wollman {\it et al},
Phys. Rev. Lett. (1993)

\bibitem{NCCO} D. H. Wu {\it et al}, Phys. Rev. Lett. 
{\bf 70}, 85 (1993)



\bibitem{vhcomm} see D. M. Newns {\it et al}, Comments on CMP
{\bf XV}, 273 (1992) for a review

\bibitem{vhtransp} D. M. Newns {\it et al}, Phys. Rev. Lett.
{\bf 73}, 1695 (1994)

\bibitem{dinf1} W. Metzner and D. Vollhardt, 
Phys. Rev. Lett. {\bf 62}, 
324 (1989), see A. Georges, G. Kotliar, W. Krauth and M. J.
Rozenberg, review article to appear in Rev. Mod. Phys.
(cond-mat 9510091), Th. Pruschke, M. Jarrell and J. K.
Freericks, preprint (to appear in Advances in Physics)


\bibitem{gkotprb} A. Georges and G. Kotliar, Phys. Rev. B
 {\bf45}, 6479 (1992).


\bibitem{iptasym}Henrik Kajueter and Gabriel Kotliar,
preprint, (cond-mat 9509152) Sept (1995 )

\bibitem{hall} Pinaki Majumdar and H. R. Krishnamurthy, preprint,
(cond-mat 9512151) Dec (1995)

\bibitem{transpcal}P. Voruganti, A. Golubentsev and
Sajeev John, Phys. Rev. B.
{\bf 45}, 13945 (1992)



\bibitem{2212pes} 
 D. S. Dessau {\it et al}, 
Phys. Rev. Lett. {\bf 71},  2781 (1993).

\bibitem{2201pes}D. M. King {\it et al}, Phys. Rev. Lett.
{\bf 73}, 3298 (1994)

\bibitem{123pes}  K. Gofron {\it
et al}, Phys. Rev. Lett. {\bf 73}, 3302 (1994)

\bibitem{nccopes} D. M. King {\it et al}, Phys. Rev. Lett. 
{\bf 70}, 3159 (1993)


\bibitem{footnote}
The asymmetry is a
consequence of the corresponding asymmetry in 
the frequency dependence of
$\Sigma(\omega)$ for a given doping.
Even in the presence of the vHs $\Sigma(\omega)$
is roughly "reflection symmetric"
for complementary $p$ and $n$ doping
and the overall "band narrowing"
is  not  very different for the two.
It is the
proximity of the renormalised vHs to $E_F$ that is
most asymmetric between $p$ and $n$ doping. The other 
features, to do with the $T$ dependence of $\Sigma(\omega=0)$
and its consequences for transport, arise from this.

\bibitem{hricecomm} R. Hlubina and T. M. Rice, Phys. Rev. B
{\bf 51}, 9253 (1995)

\bibitem{nlfs}

The gauge theories, in various versions (
SU(2) \cite{wenlee}, spin-liquid
+vHs \cite{spvhs},) 
appear to be able to reproduce the
qualitative features of this phase, although
they have difficulties in getting 
the FS right {\it far from}
half-filling. 

\bibitem{wenlee} X. G. Wen and P. A. Lee, 
Phys. Rev. Lett.  {\bf 76}, 503 (1996)

\bibitem{spvhs}L. B. Ioffe and A. J. Millis,
preprint, cond-mat 9511054


\end{references}
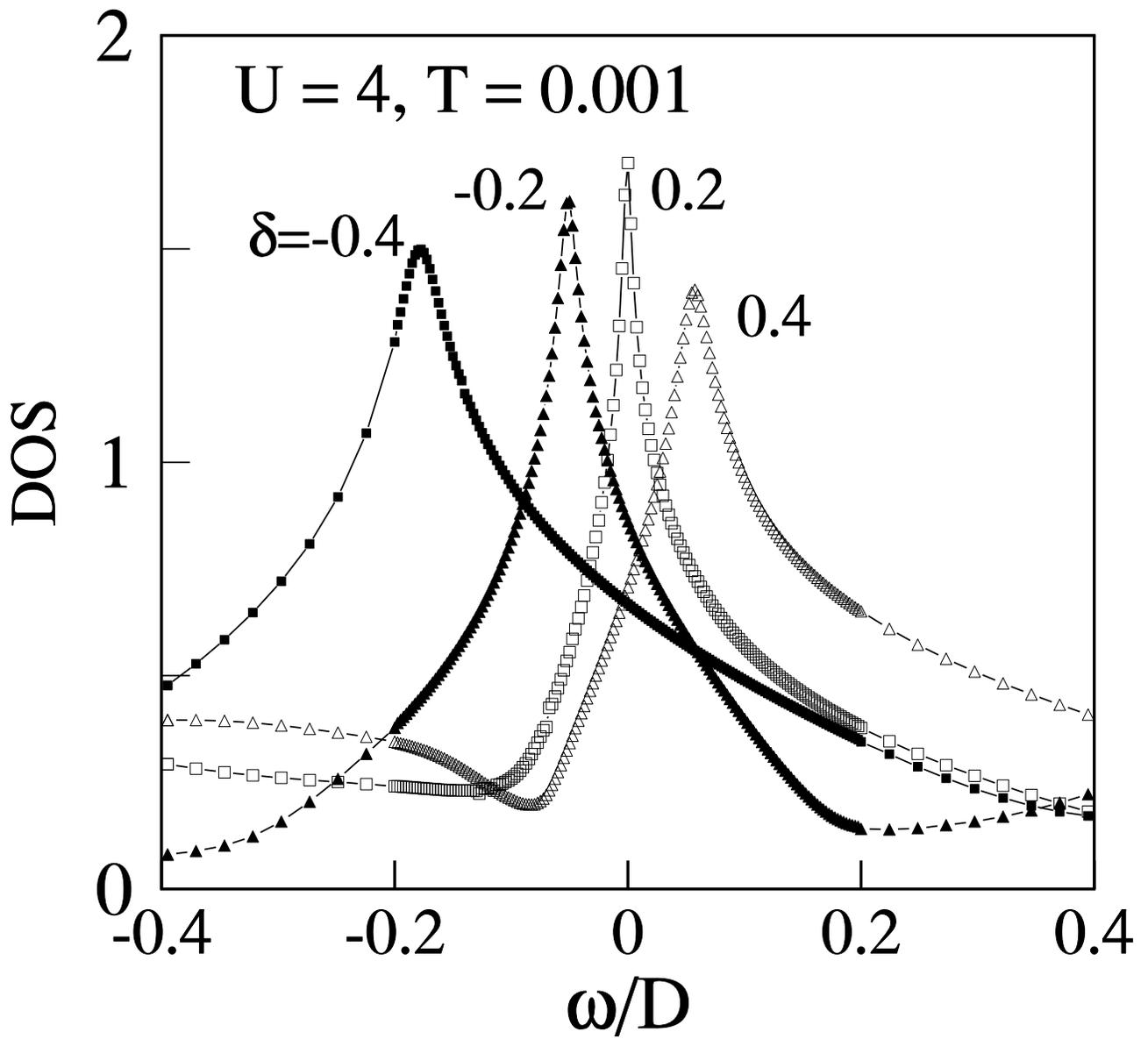
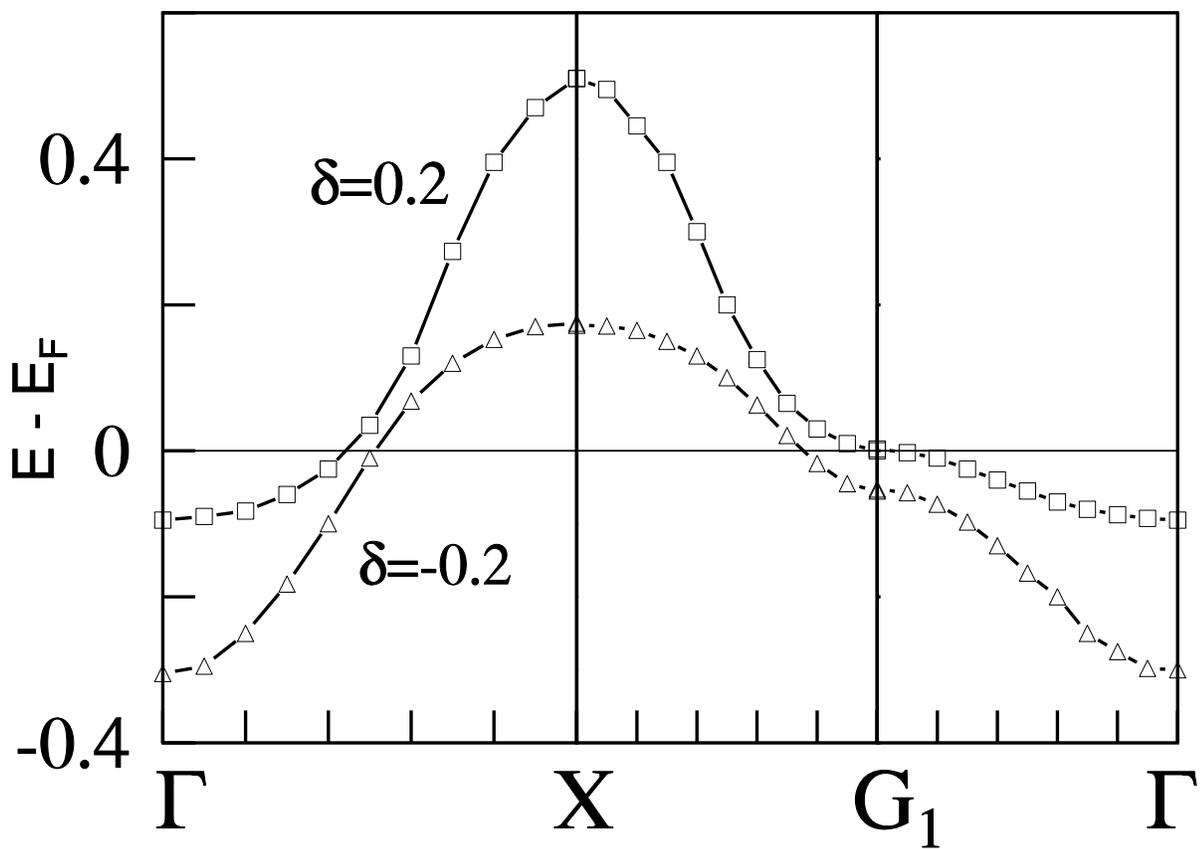
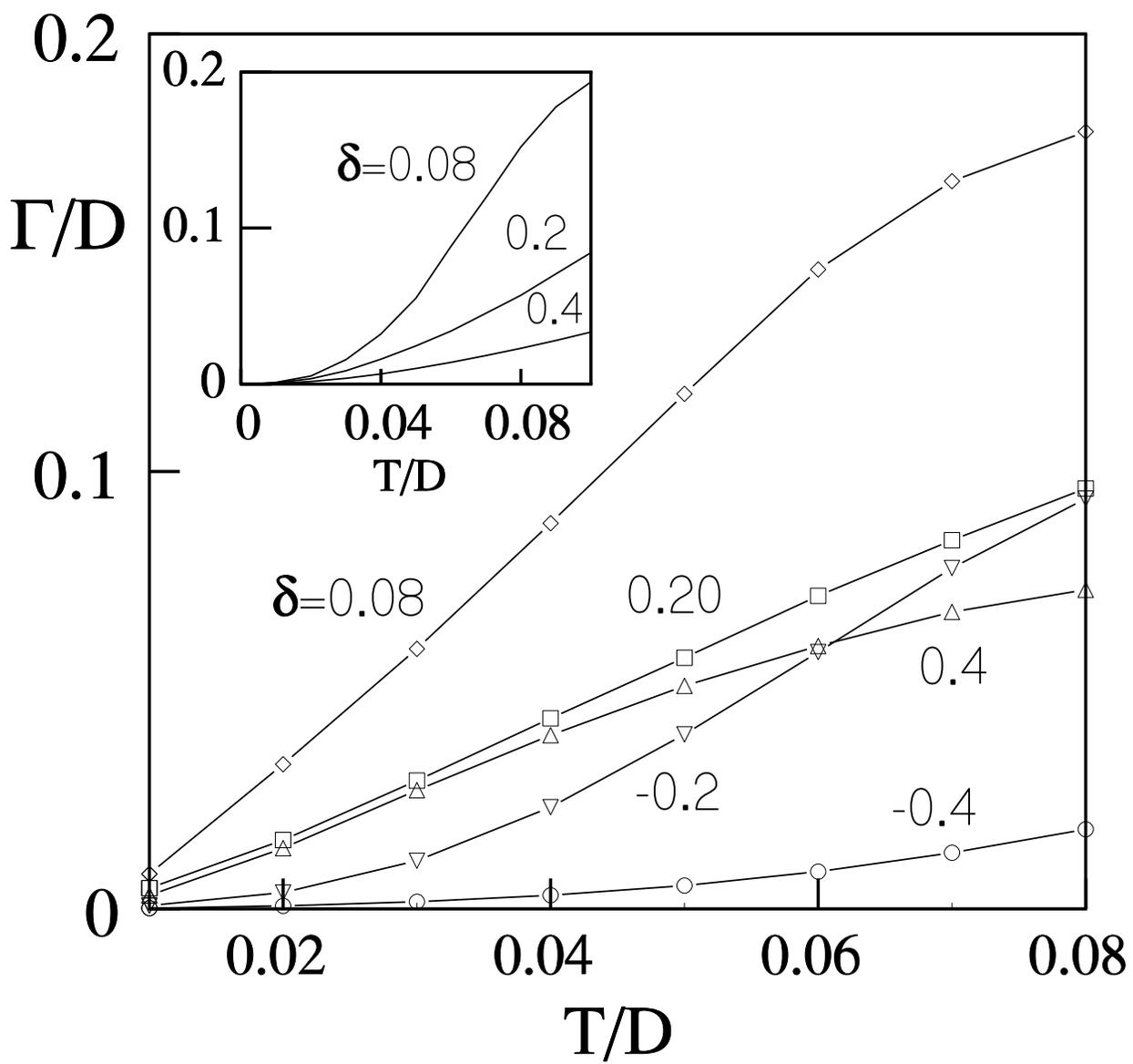
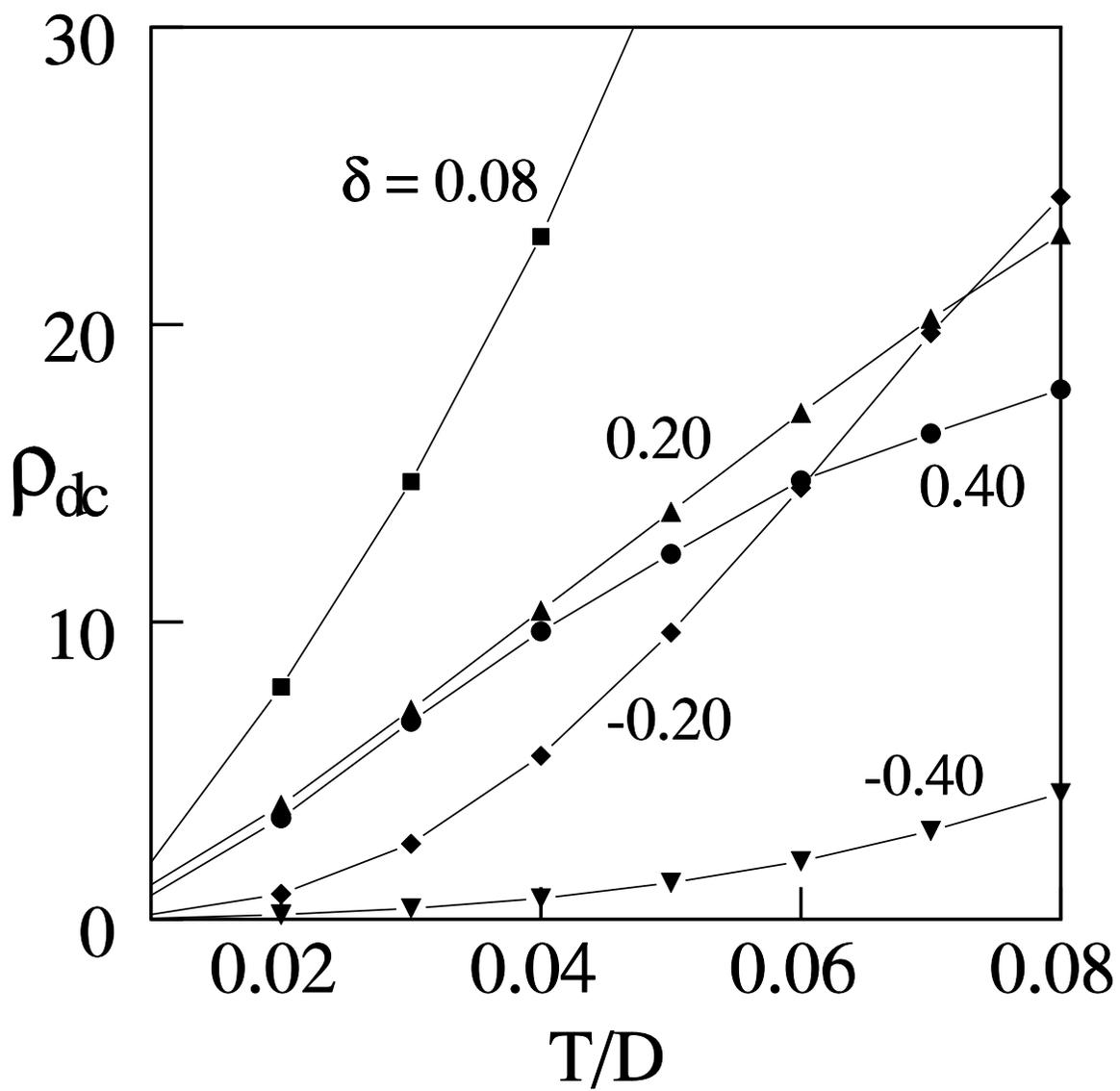
\end{document}